\documentclass{aa}  

\usepackage{graphicx}
\usepackage{txfonts}
\usepackage[dvipsnames]{xcolor}
\usepackage[version=4]{mhchem}
\begin{document}

   \title{Atmospheric characterization of HIP 67522 b with VLT/CRIRES+}

   \subtitle{VLT/CRIRES+ suggests a heavier planet and hints at deuterium fractionation}


   \author{
            A. Lavail\inst{\ref{instIRAP}} 
    \and    F. Debras\inst{\ref{instIRAP}} 
    \and    B. Klein\inst{\ref{instOXFORD}} 
    \and    E. Chabrol\inst{\ref{instLIRA}} 
    \and    S. Vinatier\inst{\ref{instLIRA}}
    \and    T. Hood\inst{\ref{instLagrange}} 
    \and    A. Masson\inst{\ref{instCAB}} 
    \and    J.  V. Seidel\inst{\ref{instLagrange}}
    \and    C. Moutou\inst{\ref{instIRAP}}  
    \and    S. Aigrain\inst{\ref{instOXFORD}} 
    \and    A. Meech\inst{\ref{instCFA}} 
    \and    O. Barragán\inst{\ref{instOXFORD},\ref{instWARWICK}}
    }

   \institute{
            Institut de Recherche en Astrophysique et Plan\'etologie, Universit\'e de Toulouse, CNRS UMR 5277, 14 avenue Edouard Belin, 31400 Toulouse, France\label{instIRAP} \\
                                \email{astro@lavail.net}
            \and Department of Physics, University of Oxford, Oxford OX1 3RH, UK\label{instOXFORD}
            \and LIRA, Observatoire de Paris, Université PSL, CNRS, Sorbonne Université, Université Paris Cité, 5 place Jules Janssen, 92195 Meudon, France\label{instLIRA}
            \and Laboratoire Lagrange, Observatoire de la C\^ote d’Azur, CNRS, Universit\'e C\^ote d’Azur, Nice, France\label{instLagrange}
            \and  Centro de Astrobiología (CAB), CSIC-INTA,  Camino Bajo del Castillo s/n, 28692, 
                                Villanueva de la Cañada, Madrid, Spain\label{instCAB}
            \and Center for Astrophysics | Harvard \& Smithsonian, 60 Garden St, Cambridge, MA 02138, USA\label{instCFA}
            \and Department of Physics, University of Warwick, Coventry CV4 7AL, UK\label{instWARWICK}
            }

   \date{Received January 30, 2026}

  \abstract
   {Young transiting exoplanets provide unique opportunities to probe planetary atmospheres during the critical early phases of evolution when atmospheric escape and contraction are most active. HIP~67522~b, a 17~Myr old hot Jupiter with an extraordinarily low bulk density ($<0.20$~g~cm$^{-3}$), represents an ideal target for high-resolution transmission spectroscopy.}
   {We aim to constrain the mass and characterize the atmospheric composition, thermal structure, and dynamics of HIP~67522~b using ground-based high-resolution near-infrared spectroscopy with VLT/CRIRES+, complementing recent JWST observations.} 
   {We obtained 92 high-resolution spectra ($R \approx 10^5$) with VLT/CRIRES+ in the K2166 band during a transit on 30 January 2025. We applied cross-correlation techniques and Bayesian nested sampling retrievals to constrain molecular abundances, temperature structure, and atmospheric dynamics.}
   {We detect H$_2$O at 20$\sigma$ and CO at 5$\sigma$, confirming the extremely extended atmosphere of this low-mass giant. A velocity offset of $-2.9 \pm 0.2$~km~s$^{-1}$ indicates day-to-night winds. The rotation velocity is constrained to $<1.8$~km~s$^{-1}$ at 3$\sigma$, consistent with tidal locking. Retrieval analysis suggests a planetary mass of $27.7^{+5.9}_{-5.5}$ Earth masses and statistically favours a two-temperature atmospheric structure with a discrete change at mbar pressures over an isothermal profile. This mass is two times larger than the mass estimated from JWST atmospheric observations and inconsistent at 3$\sigma$ hence leaving a doubt on the actual planetary density of the planet. No matter the choice of atmospheric model, we derive a supersolar C/O ratio about 1.5 times solar, and a supersolar metallicity which can be further increased if the atmosphere is cloudy, a degeneracy our data alone cannot resolve. We report a tentative 2$\sigma$ detection of HDO with an extreme enrichment factor of $\sim$1000 relative to the protosolar D/H ratio. If confirmed, this would be the first detection of deuterium in an exoplanet atmosphere and would require intense escape rate to be explained.}
   {}

   \keywords{ Techniques: spectroscopic --
                Planets and satellites: atmospheres --
                Planets and satellites: individual: HIP 67522
               }

   \maketitle

\section{Introduction}

Young transiting exoplanets ($< 100$~Myr) represent crucial laboratories for understanding planetary formation as they allow direct observation of planets still undergoing contraction, cooling, and atmospheric escape. These systems provide empirical benchmarks for theories of evolution during the critical first hundred million years.
Probing the atmospheres of these young planets provides a unique window into the early stages of planetary evolution, when key processes such as atmospheric escape and orbital migration are most active \citep{Owen2019,Baruteau2016}. The atmospheric composition, temperature structure, and dynamics of young planets carry the imprint of their formation history. However, characterizing young planetary atmospheres remains challenging. Only a handful of stars younger than 100~Myr are known to host transiting planets, and their youth manifests as high magnetic activity---frequent flares and starspots---that complicates the determination of planetary masses and atmospheric properties \citep[e.g.][]{Rackham2018}. 
The youngest confirmed transiting systems include TIDYE-1b ($\sim$3~Myr; \citealt{Barber2024}), a Jupiter-sized planet still embedded in its natal disk, TOI-1227~b ($\sim$8~Myr; \citealt{Mann2022}), K2-33~b ($\sim$5--10~Myr; \citealt{Mann2016}), AU~Mic~b and c ($\sim$20--24~Myr; \citealt{Plavchan2020}), the remarkable four-planet V1298~Tau system ($\sim$23~Myr; \citealt{David2019}) and the puzzling planet HIP~67522~b ($\sim$17~Myr; \citealt{Rizzuto2020}).

HIP~67522 (=HD~120411) is a 17~Myr old early G star in the Sco-Cen association at a distance of 127~pc \citep{Rizzuto2020}. It hosts what was thought to be the youngest known transiting hot Jupiter because of its radius of $\sim 10$~R$_\oplus$ , HIP~67522\,b,  with an orbital period of 6.96~days \citep{Rizzuto2020}, and a second outer planet at 14.33~days \citep{Barber2024}. \citet{Heitzmann2021} constrained the projected obliquity of the planet to be $|\lambda| = 5.8^{+2.80}_{-5.7}\,^\circ$. Recent JWST/NIRSpec observations have revealed that HIP~67522\,b has an extraordinarily low bulk density ($< 0.10$~g~cm$^{-3}$) and suggested a mass of only $13.8 \pm 1.0$~M$_\oplus$ \citep{Thao2024}. This extremely low density translates into an atmospheric scale height comparable to the planetary radius itself, making HIP~67522\,b one of the most favorable targets for atmospheric characterization. The JWST transmission spectrum showed strong detections of H$_2$O and CO$_2$ ($\ge 7\sigma$), with modest detection of CO (3.5$\sigma$) and SO$_2$ (2 $\sigma$), indicating a metal-enriched atmosphere with 3--10$\times$ solar metallicity and a near-solar to sub-solar C/O ratio \citep{Thao2024}.

Recent photometric monitoring campaigns with TESS and CHEOPS have revealed a remarkable property of this system: the orbital period of HIP~67522\,b exhibits significant clustering of stellar flares, with an approximately nine-times higher flare rate occurring in the 20\% of the orbit immediately following planetary transit \citep{Ilin2025a}. This represents a robust detection of planet-induced stellar flaring. Follow-up radio observations at 1.1--3.1~GHz with ATCA confirmed strong radio activity consistent with coronal emission of cool dwarfs but did not detect the expected electron cyclotron maser emission (ECME) signature of magnetic star-planet interaction, placing upper limits of $<0.7$\% on the conversion efficiency of interaction power into radio waves \citep{Ilin2025b}.

Ground-based high-resolution spectroscopy (HRS) in the near-infrared provides a complementary approach to space-based observations, with the ability to simultaneously constrain atmospheric composition, temperature structure, and dynamics through the Doppler shifts and line broadening of molecular absorption features \citep{Snellen2010,Brogi2016}. In this paper, we present VLT/CRIRES+ high-resolution transmission spectroscopy of HIP~67522\,b in the K band (centred at 2166~nm), obtained during a transit on 30 January 2025. Our observations and data reduction are presented in Sect.~\ref{sec:observations_and_data_reduction}, the cross-correlation analysis and atmospheric retrievals are detailed in Sect.~\ref{sec:cross-correlation-bayesian-exploration} and we discuss our results in Sect.~\ref{sec:discussions-conclusions}.

\section{Observations and data reduction}
\label{sec:observations_and_data_reduction}

\subsection{Description of the observations}
We observed HIP 67522 during a transit of the exoplanet HIP 67522~b. We used the high-resolution near-infrared spectrograph CRIRES+ \citep{Dorn2023} mounted at the 8-metre Unit Telescope 3 of the Very Large Telescope on the 30th January 2025 (ESO Programme 114.28HP.001, PI: Klein). Our observations started at 04:28~UT, ended at 09:18~UT and consisted of 92 consecutive spectra each taken with a detector integration time of 180 seconds. The observations were carried out in nodding mode, meaning that the target was alternatively positioned on two distinct positions (A and B) on the slit. The nodding procedure facilitates the data reduction and the removal of sky and instrumental background. 

\begin{figure}
\centering
\includegraphics[width=\columnwidth]{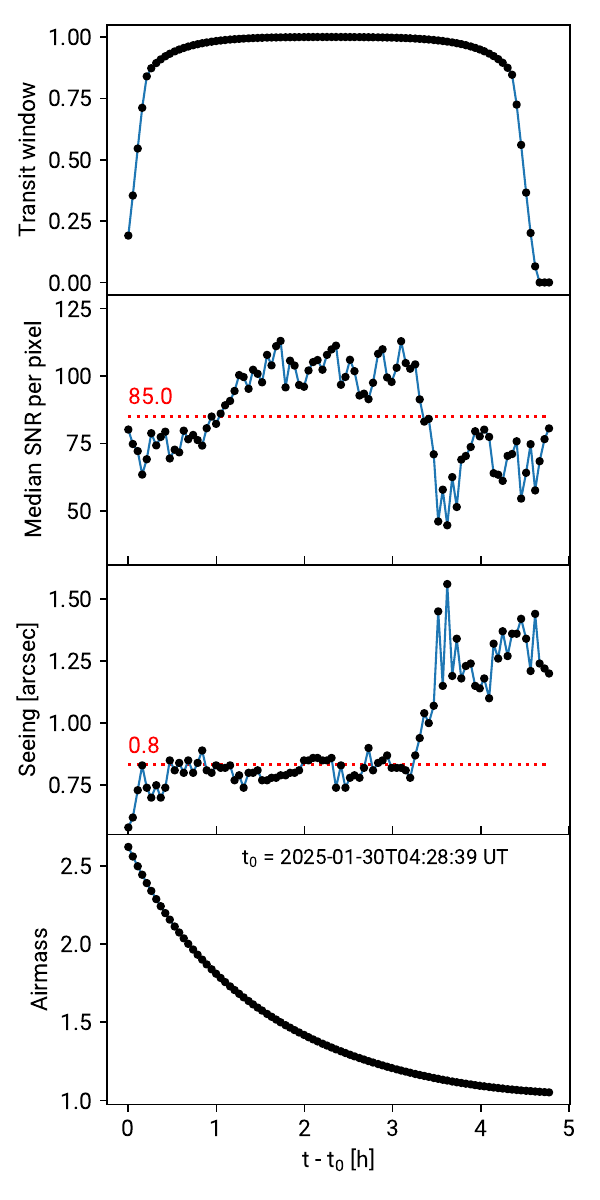}
\caption{Evolution of the transit window, the median SNR per exposure, seeing, and airmass as a function of time from the first exposure. The median SNR and median seeing for the full time-series are indicated in red on the second and third panel respectively. The UT time at the start of the first exposure $t_0$ is indicated on the fourth panel.}
\label{figure:obslog}
\end{figure}

We used CRIRES+ in the K2166 wavelength setting covering the wavelength range between 1990 and 2472~nm over 6 spectral orders -- note that there are wavelength gaps between the orders. Metrology was activated during the acquisition to improve the accuracy of the wavelength solution. The 0.2" slit was set up yielding the nominal resolving power of $R \approx 10^5$. This dataset does not suffer from the known super-resolution issue  \citep[e.g.][]{2025nortmann} affecting CRIRES+ spectra for which the point spread function (PSF) is smaller than the slit width. In this case, we did not use the adaptive optics system, and the PSF stayed larger than the slit throughout the observations. The seeing increased drastically in the last third of the transit and decreased the signal-to-noise ratio (SNR) accordingly despite an improving airmass (Fig.~\ref{figure:obslog}).

\subsection{Data reduction}
\label{ssec:reduc1}
The data were reduced with the CRIRES+ data reduction system (DRS) \verb!cr2res! available from the ESO website\footnote{\url{https://www.eso.org/sci/software/pipelines/cr2res/}}. In a first step, the raw calibrations (flat fields, darks, Fabry-Perot etalon, Uranium-Neon lamp) taken as part of the VLT daily calibration routine and associated to our science data were reduced using the standard calibration cascade, as described in the CRIRES+ pipeline user manual. This step yields reduced calibrations : (1) normalized flat fields characterising pixel-to-pixel sensitivity variations, (2) a bad pixel mask, and (3) a so-called tracewave file containing the locations of the spectral orders on the detectors, the slit curvature, and the wavelength solution. In a second step, reduced calibrations are applied to pairs of spectra taken in nodding positions A and B. Finally, 1D spectra are extracted with the \verb!cr2res_obs_nodding! recipe. This step provides the time-series of 92 1D wavelength-calibrated spectra.

After the DRS reduction, we apply \verb!molecfit! on the mean A and B spectra to further improve the wavelength solutions for the A and B nodding position. We then interpolate the B spectrum into the wavelength solution of the A spectrum. We also compute the BJD TDB (Barycentric Julian Date in the Temps Dynamique Barycentrique time scale) timestamp at midexposure for each spectra as well as the barycentric velocity correction (BERV) using {\tt barycorrpy} \citep{Kanodia2018}. 

\subsection{Additional post-processing steps}
\label{ssec:reduc2}
After this initial data reduction, the data are shaped so that we apply the ATMOSPHERIX pipeline\footnote{ \url{https://github.com/baptklein/ATMOSPHERIX_DATA_RED}} to reduce transmission spectroscopy data and extract planetary signal \citep{Klein2024,Debras2024}. 
The pipeline is made up of different steps, summarised as follows.
\begin{enumerate}
   
    \item The spectra are all aligned in the stellar rest frame, from which a master-out spectrum I$_\mathrm{ref}$ is created usually by averaging out-of-transit exposures. In our case, because we have too few baseline observations, the master spectrum encompasses all exposures. This master spectrum is then moved back to the geocentric frame and linearly matched in flux to each of the observed spectra, which are each divided by its best-fitting solution. The resulting spectra are then divided by a second master spectrum,  now in the Earth rest frame, to provide an additional correction of the tellurics. 
    
    \item A normalisation of each of the resulting spectra is then performed using an estimate of the noise-free continuum, calculated using a rolling mean window of 150 pixel. This is followed by a 5$\sigma$ clipping being applied to remove outliers. This two-step process is then repeated until there are no more outliers flagged in the data.

    \item As some pixels with high temporal variance might remain, we calculated the variance for each pixel and applied an iterative parabolic fit to the pixel variance distribution. We considered pixels further than 5$\sigma$ from the fit as outliers and masked them out for the rest of the data-reduction process.

    \item Finally, we further correct remaining correlated noise through the use of a principal component analysis (PCA). In this study, we removed 6 principal components for each CRIRES+ order, as fewer components did not allow to correct the tellurics properly and we could observe their influence in the K$_p$-$v_{\textrm{sys}}$ map up to 5 principal components. 
\end{enumerate}

Applying a PCA to the data also affects the planetary signal, degrading it.  To take into account this degradation, we implemented the method of \cite{gibson2022}. For this, during the data-reduction process, we keep a matrix U of the removed eigenvectors for each order which by construction is associated with correlated noise. These are subsequently used to prepare the synthetic spectra for analysing the data to degrade them coherently with the real signal. 
We have shown that using this method in our ATMOSPHERIX pipeline allows to retrieve unbiased physical parameters of the planetary atmosphere using synthetic data in \citet{Klein2024}, excluding statistical sources of biases discussed in \citet{Debras2024}.

\section{Cross-correlation maps and Bayesian exploration}
\label{sec:cross-correlation-bayesian-exploration}
\subsection{Detection of H$_2$O and CO through correlation}
\begin{figure}
\centering
\includegraphics[width=\columnwidth]{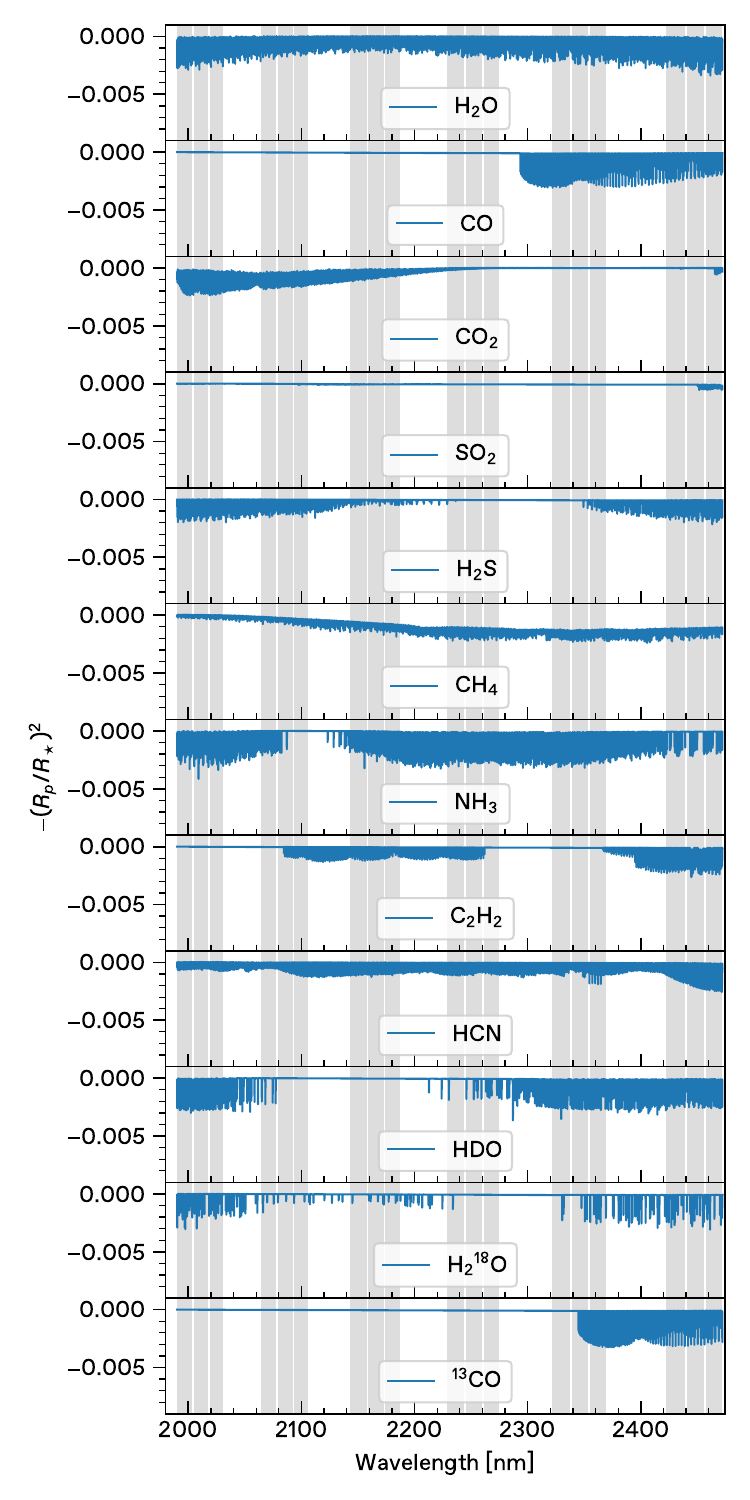}
\caption{Model isothermal spectra for each of the chemical species we investigated in the atmosphere of HIP 67522~b. The gray shaded areas represent the wavelength coverage of CRIRES+ with six spectral orders each spread over three detectors.}
\label{figure:synth-spec}
\end{figure}
Once the data is reduced, we searched for molecular absorption in the atmosphere of HIP 67522 b starting by species detected by JWST (H$_2$O, CO, CO$_2$ and H$_2$S) and extending to other plausible species in the atmosphere, based on thermo-chemical models (CH$_4$, NH$_3$, C$_2$H$_2$, and HCN). The physical values used for our models are gathered in Table \ref{tab:sysparams}.  Note that the mass is almost two times the mass derived from JWST observations in \citet{Thao2024}. This is explained and justified in Sect.~\ref{ssec:simple_retrieval}.

\begin{table*}
    \centering
    \caption{Adopted HIP 67522 system parameters. We separated the parameters we retrieved with the isothermal or two-temperature models (see text).The other parameters are common in both models. The metallicity and C/O ratio are only determined with H$_2$O, CO and HDO.}
    \label{tab:sysparams}
    {\renewcommand{\arraystretch}{1.42}%
    \begin{tabular}{lccr} 
        \hline
        Stellar parameters & \multicolumn{2}{c}{ Value} & Reference\\
        \hline
        Mass (M$_\odot$) &  \multicolumn{2}{c}{1.22} & \cite{Rizzuto2020} \\
        Radius (R$_\odot$) &  \multicolumn{2}{c}{1.38} & \cite{Rizzuto2020} \\
        Effective temperature (K) & \multicolumn{2}{c}{5675} & \cite{Rizzuto2020} \\
        Systemic velocity (km.s$^{-1}$) &  \multicolumn{2}{c}{7.4} & \cite{Rizzuto2020} \\
        \hline
        Planetary parameters &   \multicolumn{2}{c}{Value} & Reference\\
        &   Isothermal & Two-temperature & \\

        \hline
        Epoch of transit (T$_0$) &   \multicolumn{2}{c}{2458604.02376}  & \cite{Barber2024} \\
        Orbital Period (days) &    \multicolumn{2}{c}{6.9594731}  & \cite{Barber2024} \\
        Mass (M$_J$) & 0.087 & 0.11 & This work \\
        Mass (M$_\mathrm{Earth}$) & 27.7 & 34.5 & This work \\
        1 $\sigma$ Mass uncertainty (M$_\mathrm{Earth}$) & $^{+5.9}_{-5.5}$ & 7.7 & This work \\
        Planet-star radius ratio &  \multicolumn{2}{c}{0.0668} & \cite{Rizzuto2020} \\
        Uncertainty on planet-star radius ratio $\sigma_\mathrm{TESS}$ &  \multicolumn{2}{c}{0.001} &\cite{Rizzuto2020} \\
        Radius (R$_J$) &  \multicolumn{2}{c}{0.892} & \cite{Rizzuto2020}  \\
        g (m.s$^{-2}$) & 2.62 & 3.27 & This work \\
        Planet RV semi-amplitude (km.s$^{-1}$) &  \multicolumn{2}{c}{120} & \cite{Barber2024} \\
        Semi-major axis (au) &  \multicolumn{2}{c}{0.075} & \cite{Barber2024} \\
        Inclination (deg) &  \multicolumn{2}{c}{89.3}& \cite{Rizzuto2020} \\
        Eccentricity &  \multicolumn{2}{c}{0.0} & Fixed \\
        Transit duration (h) &  \multicolumn{2}{c}{4.82} & \citet{Rizzuto2020} \\
        Equilibrium Temperature (K) &  \multicolumn{2}{c}{1100} & \citet{Rizzuto2020} \\
        Metallicity ([C+O/H]) & $+0.6^{+0.4}_{-0.4}$ & $+1.9^{+0.4}_{-0.4}$ & This work \\
        C/O ratio & $0.87^{+0.05}_{-0.09}$& $0.92^{+0.04}_{-0.07}$ & This work \\
        \hline
    \end{tabular}}
\end{table*}

We created isothermal models of the planet atmosphere (Fig.~\ref{figure:synth-spec}) using {\tt petitRADTRANS} \citep{molliere2019,blain2024} containing only one of each of these species with a mass mixing ratio of 10$^{-3}$ and correlated these models with our reduced data. In {\tt petitRADTRANS}, gravity is not assumed constant across the atmosphere. The input value defines the gravity at a given reference pressure and radius and the actual gravity in the codes varies with $r^2 $ (where $r$ is the planet’s radial coordinate) from this value, as is expected from Poisson equation and assuming that the atmosphere has negligible influence on the mass (hence neglecting self gravity). We thereafter computed the so-called K$_p$-$v_{\textrm{sys}}$ map where a detection is validated when the maximum of correlation is significantly larger than the noise level, at the expected orbital semi-amplitude of the planet and at physically sound Doppler shifts. The isothermal model spectra for each species are shown in Fig.~\ref{figure:synth-spec}. 

\begin{figure*}
\centering
\includegraphics[width=\textwidth]{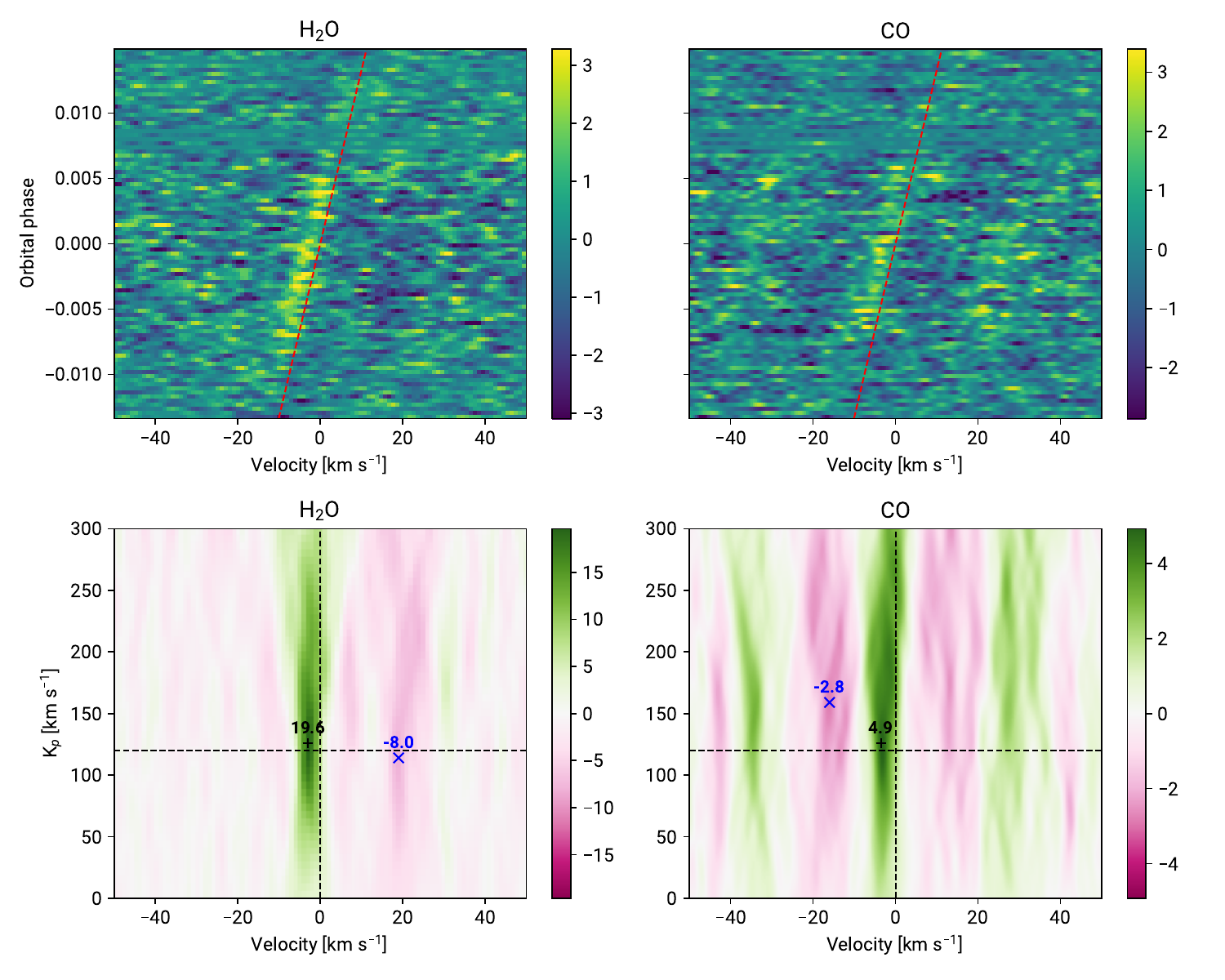}
\caption{Top panel: Phase-$v_\textrm{sys}$ maps for the detected species H$_2$O (left) and CO (right) in the stellar reference frame. The red dashed line indicate the predicted velocity trail of the planet. The colorbar are expressed in standard deviations away from both planet and tellurics signals (excluding the square in the K$_p$-V$_\mathrm{sys}$ map defined by K$_p \in [0,300]$ and V$_\mathrm{sys} \in [-10,30]$). Bottom panel: K$_p$-$v_{\textrm{sys}}$ map for H$_2$O (left) and CO (right), the vertical dashed line indicate stellar reference frame, the horizontal dashed line indicated K$_p = 120 $~km s$^{-1}$. The maximum value of the map is indicated by the + symbol in black with the associated value, the minimum is indicated by the blue cross.}
\label{figure:phase-vsys}
\end{figure*}

Through this method, we detected H$_2$O at 20$\sigma$ and CO at 5$\sigma$ (Fig.~\ref{figure:phase-vsys}). Here, $\sigma$ is evaluated from the standard deviation of the noise in the K$_p$-$v_{\textrm{sys}}$ map (selecting the area of the map where planetary signals and telluric residuals are absent). Under this definition, a 20$\sigma$ detection corresponds to a correlation signal 20 times stronger than the noise standard deviation. We did not detect any of the other available species through this method, some of them like SO$_2$ being almost impossible to detect in the K-band due to their lack of strong absorption lines. The extremely high level of detection of water and carbon monoxide compared to usual detections in near infrared high resolution spectroscopy confirms an extremely extended atmosphere, hence a low-mass planet. 

The maximum of detection occurs at a velocity of $-3$~km.s$^{-1}$ relatively to the systemic velocity, which indicates day-to-night winds on the limbs of the planet. We adopted a systemic velocity of $v_\textrm{sys} = 7.41 \pm 0.25$~km s$^{-1}$ \citep{Rizzuto2020}. The exact value of these winds is difficult to measure, as the star is a fast rotator. Nonetheless, regarding the results by different instruments at differents epoch in \citet[][Table 2]{Rizzuto2020}, the systemic velocity of the star appears rather superior to the value we chose, and hence the wind speed is more likely underestimated. 

\subsection{Bayesian exploration through nested sampling}
\subsubsection{Isothermal models}
\label{ssec:simple_retrieval}
In order to constrain further the atmospheric composition and thermodynamics, we then ran nested sampling retrievals, based on the {\tt PyMultiNest} code \citep{Feroz2008,Feroz2009,buchner2014}. For all the retrievals presented in this work, we ran {\tt PyMultiNest} with a tolerance on the evidence of 0.5 and a sampling efficiency of 0.8 which are default values. We used 512 live points when having six or less free parameters, and 1024 live points otherwise. When we calculated Bayes factors (see section \ref{ssec:Bayes}) we also used 1024 live points.

We first used isothermal models and started by looking for an independent estimation from JWST on the mass of the planet. Note that the mass is not directly a parameter of the retrieval, as the physical quantity that impacts the atmosphere scale height is gravity. We therefore define here "mass" as: 

\begin{equation}
    M = \dfrac{gR^2_\mathrm{TESS}}{\mathcal{G}}
\end{equation}

\noindent where $g$ is the gravity that we retrieve, $\mathcal{G}$ is the gravitational constant and R$_{\mathrm{TESS}} = 0.892$ R$_\mathrm{J}$ the best fit radius from TESS observations with R$_\mathrm{J}$ the Jupiter mean radius and assuming a stellar radius of 1.38 R$_\odot$ with R$_\odot$ the solar radius. 

Because the radius of the planet is extremely dependent on the wavelength, as the atmosphere is almost the size of the planet, we included the TESS radius in our retrievals by defining a TESS likelihood \citep[where the constant terms are neglected as in ][]{Brogi2019}: 

\begin{equation}
\ln \mathcal{L}_{\mathrm{TESS}} = -\dfrac{1}{2} \dfrac{((R_\mathrm{p}/R_\star)_\mathrm{TESS}  -(R_\mathrm{p,retrieved}/R_\star))^2}{\sigma_\mathrm{TESS}^2}
 \end{equation}
 
\noindent where R$_\mathrm{p}$ is the planetary radius, R$_\star$ the stellar radius, TESS subscript indicates the TESS observations, "retrieved" indicates the retrieved value by our nested sampling algorithm. To calculate the TESS radius in our observations, we integrated the radius obtained in our isothermal models multiplied by the instrumental response of TESS in its wavelength range of observation.

Contrary to low resolution observations, we are not able to directly compare our best-fit  model to the data, as they are drowned under the noise. We therefore follow \citet{Brogi2019} and include a scaling parameter $a$ in the high resolution spectroscopy likelihood:
 
 \begin{equation}
\ln \mathcal{L}_{\mathrm{HRS}} = -\dfrac{1}{2} \sum_{n}{\dfrac{(\mathrm{data}[n]  -a * \mathrm{model}[n])^2}{\sigma[n]^2}}
 \end{equation}
 where data[n] is the data at each pixel, model[n] the value of the retrieved model during the bayesian exploration at the same pixel and $\sigma[n]$ the observational uncertainty at each pixel, defined by the standard deviation in time of each pixel as in \citet{Klein2024}. Our models can be considered correct when the retrieved $a$ distribution is centered around 1. As in \citet{Brogi2019}, our total likelihood is simply:
 \begin{equation}
\ln \mathcal{L}_{\mathrm{tot}} =\ln \mathcal{L}_{\mathrm{TESS}}+\ln \mathcal{L}_{\mathrm{HRS}}
 \end{equation}

Note that we do not include a $\beta$ parameters to scale the noise as we follow the method of \citet{Gibson2020} that derives the likelihood with respect to this parameter to minimize its value. 

The parameters of our isothermal retrieval are therefore K$_p$, V$_\mathrm{sys}$, isothermal temperature, water and carbon monoxide mass mixing ratios (MMR), a corrective factor for gravity (dg), a corrective factor for radius (dR), the reference pressure $P_0$\footnote{the reference pressure is the pressure where the gravity is equal to the input gravity, and hence the radius is equal to the one used in the code (here 0.892 Jupiter radii)}, and the $a$ factor in the likelihood.
Firstly, including the end of the transit, where the seeing gets two times larger than during the beginning of the night (see Fig. \ref{figure:obslog}), biases our retrievals towards unrealistic values of K$_p$ excluding the expected planetary value at 2$\sigma$. However, when considering only the low-seeing part of the observation, we do retrieve that the posterior for orbital velocity is centered on the expected planetary value. In the rest of the bayesian exploration, we therefore limit ourselves to the exposures before the drastic increase of the seeing.

Regarding mass constraints, we obtain that dg is directly correlated with the $a$ value, which is expected as the size of the atmosphere is directly proportional to gravity  \citep[see notably the discussion in][]{Debras2024}. We therefore only kept dg and fixed $a=1$ in the likelihood. This gives us a retrieved mass of $27.7^{+5.9}_{-5.5}$ Earth masses for the planet (1$\sigma$ errorbars) which is about two times the mass retrieved by \citet{Thao2024}. We are therefore inconsistent with their mean retrieved mass at 3$\sigma$. We discuss this discrepancy in Sect.~\ref{ssec:mass}. 
 
With these isothermal models we confirmed the detection of H$_2$O and CO and were able to put constraints on the maximum MMR of other species as detailed in Table \ref{tab:molecules}. We also confirmed a wind speed of $-2.9 \pm 0.2$~km.s$^{-1}$, consistent with our correlation maps. 

\begin{table*}
    \centering
    \caption{Molecular abundances from isothermal and Two-temperature retrievals. We considered that the 1$\sigma$ abundances were more constrained than upper limits when (i) the posterior showed a peak, and not a flat uniform distribution and (ii) the lower 1$\sigma$ abundance limit was superior to -9. The HDO distribution has the highest peak of all molecules (excepting H$_2$O and CO), and the only one with a 2$\sigma$ lower bound $>$ -9 no matter the temperature model.  }
    \label{tab:molecules}
    {\renewcommand{\arraystretch}{1.42}%
    \begin{tabular}{l|cc|cr} 
        \hline
        Molecules & \multicolumn{2}{c|}{Isothermal} & \multicolumn{2}{c}{Two temperature}\\
        & 1$\sigma$  ($\log_{10}$(MMR)) & 3$\sigma$ ($\log_{10}$(MMR)) & 1$\sigma$  ($\log_{10}$(MMR)) & 3$\sigma$ ($\log_{10}$(MMR)) \\
        \hline
        H$_2$O &  $\in$ [-3.6,-3.1] &   $\in$ [-4.1,-2.4] & $\in$ [-2.5,-2.] & $\in$ [-2.9,-1.5]\\
        CO & $\in$ [-3.2,-1.4] &  $\in$ [-3.6,-1.] & $\in$ [-1.6, -0.76]& $\in$ [-2.45, -0.03]\\
        CO$_2$ & $<$ -5.2 & $<$ -3.1 & $<$ -4.7 & $<$ -2.5 \\
        SO$_2$ & $\in$ [-8.5,-2.7] & $<$-1.  & $\in$ [-8.4,-2.7]  &  $<$-1.\\
        H$_2$S & $<$ -4.7 & $<$ -2.9 &$<$-3.9 & $<$-2.\\
        CH$_4$ & $<$ -6.4 & $<$ -4.8 &$<$-6.0 & $<$-4.4\\
        NH$_3$ & $\in$ [-7.2, -5.2] & $<$ -4.3 & $\in$ [-6.8,-4.3] & $<$-3.6\\
        HCN & $\in$ [-7.7,-3.6] & $<$-2.8 & $\in$ [-8.2,-3.0]&  $<$-2.\\
        C$_2$H$_2$ & $<$ -5.6 & $<$-3.9&   $<$-4.8. &  $<$-3.0 \\ 
        HDO & $\in$ [-5.7,-4.4]  & $<$ -3.3  & $\in$ [-4.5,-3.4]& $<$ -2.5 \\
        \ce{H2^{18}O}  & $<$ -5.6 & $<$-3.5 &  $<$-4.9 &  $<$-2.6 \\
        $^{13}$CO & $<$  -4.6 & $<$ -2.7 &  $<$-3.5 &  $<$-1.6\\
        \hline
        \hline
    \end{tabular}}
\end{table*}

With our isothermal retrieval, only considering water and CO we obtain a C/O ratio of $0.87^{+0.05}_{-0.09}$ , which is about 1.5 times solar. Note that we are limited to C/O ratio inferior or equal to 1 as we only retrieve water and CO, and that might bias our retrieved ratio towards lower values \citep[see discussion in][]{Hood2024}. We further derive a metallicity [C+O/H] of $+0.6^{+0.4}_{-0.4}$. The C/O ratio is inconsistent with \citet{Thao2024}, and we discuss it further in Section \ref{sec:discussions-conclusions}. 

We included the possibility for an opaque cloud deck, and we obtain the usual result that clouds are degenerated with metallicity (without affecting the C/O ratio): a higher metallicity leads to an absorption that can occur above the cloud deck, provided the clouds are deeper than 0.1 mbar. As discussed in Sect. \ref{sec:discussions-conclusions}, we are not able to resolve this degeneracy, and would need additional observations at different wavelength ranges.

\begin{figure*}
\centering
\includegraphics[width=\textwidth]{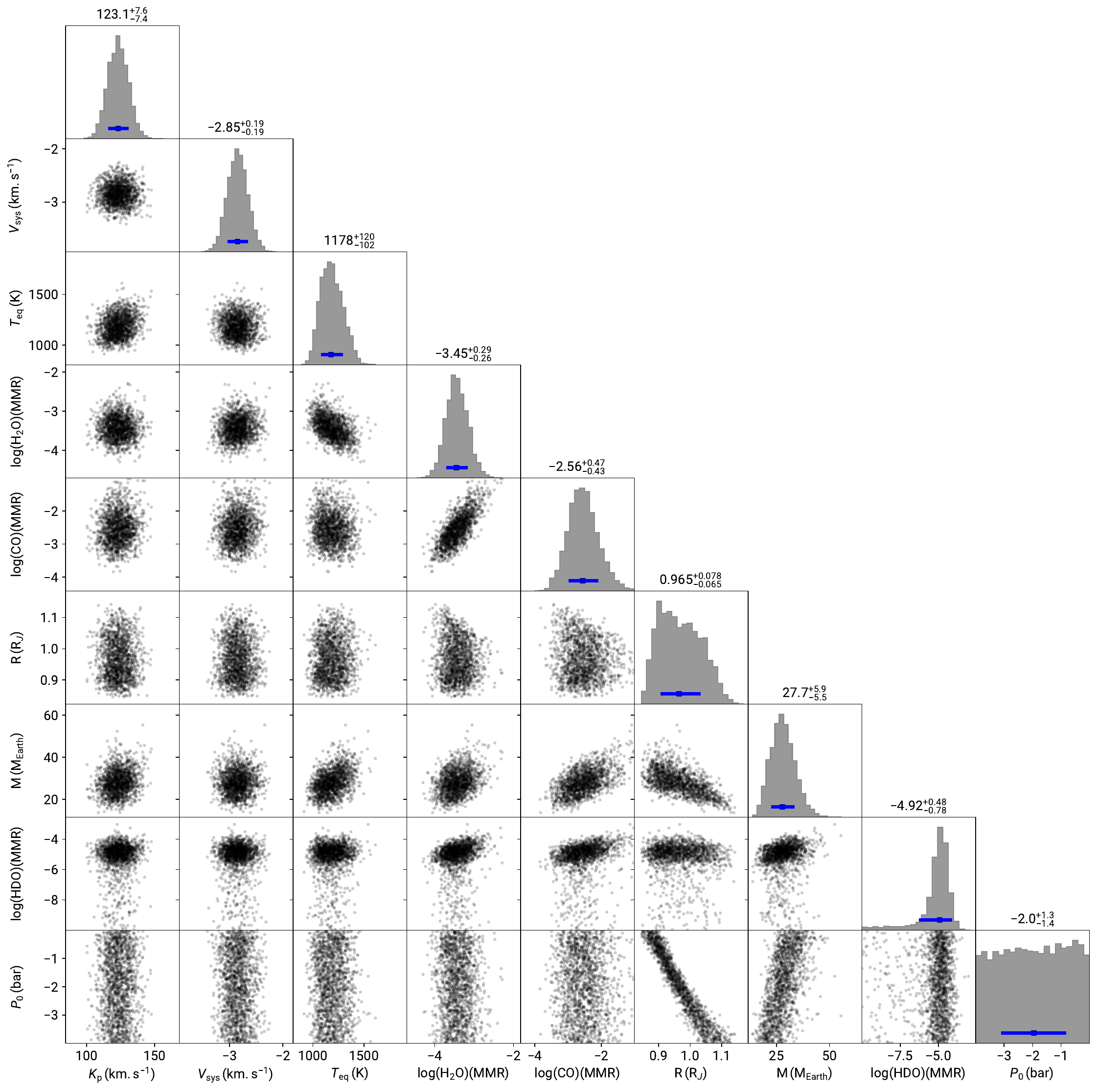}
\caption{Corner plot of the posterior distribution of our isothermal models. }
\label{figure:marg_P0}
\end{figure*}

We also put constraints on the rotation speed of the planet using the efficient rotational kernel detailed in the appendix of \citet{Klein2024}. We only obtained an upper limit: the data favour a rotation speed at the equator inferior to 0.9~km.s$^{-1}$ at 1$\sigma$ and 1.8~km.s$^{-1}$ at 3$\sigma$. This is consistent with tidal locking and excludes Jupiter-like rotation speed (around 10~km.s$^{-1}$ at the equator). Given the fact that the atmosphere is about the size of the planet, tidal locking would yield a rotation speed of 0.4~km.s$^{-1}$ at the equator around 1~bar and up to 0.8~km.s$^{-1}$ at 0.1~mbar pressure. 
In Fig.~\ref{figure:marg_P0}, we show the posterior densities of the parameters of our isothermal model including HDO (see Sect. \ref{ssec: HDO}), that shows visually the results as discussed here and in the rest of the paper. 

\subsubsection{Two temperature model}
\label{ssec:complex-retrievals}

Because of the extremely high signal to noise ratio obtained in cross correlation maps, we decided to go a step further in the usual bayesian analysis and used a model with a discrete change in temperature in the atmosphere, that we called the "two-temperature model". In a first step, we let the pressure at which this temperature change occur as a free parameter in the retrieval. We obtained that the retrieval favours a much larger deep temperature with a change between both temperatures favoured around a few mbar (2400$\pm 400$K in the interior, 1100 $\pm 100$K above). The evidence ratio between this two-temperature model and the isothermal model is of 3 (see Section \ref{sec:discussions-conclusions}): the two-temperature model is not drastically favoured compared to an isothermal.

\begin{figure}
\centering
\includegraphics[width=\columnwidth]{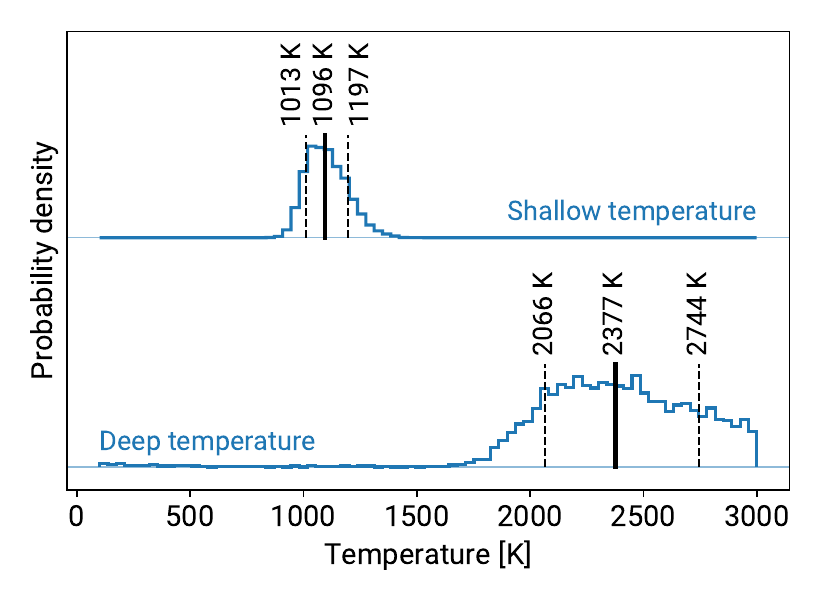}
\caption{Posterior distribution for the two-temperature model where the pressure change is fixed at 3mbar.}
\label{figure:T2pts}
\end{figure}

However, when we fix the pressure at which the temperature change occur to the best retrieved pressure (3 mbar), we obtain a similar strong difference in temperature between the two part of the atmosphere but a Bayes factor of 20: the data strongly favour a two temperature model with a change at mbar pressures. We therefore re-estimated the metallicity and C/O ratio in the two-temperature model and display the result as an additional column in Table \ref{tab:sysparams}, as well as the abundances in Table \ref{tab:molecules}. We also show the retrieved temperatures in Figure \ref{figure:T2pts}. The corner plot of the posterior distribution for the two-temperature model is shown in Fig.~\ref{figure:marg2T}.

\subsubsection{Tentative detection of HDO}
\label{ssec: HDO}

We searched for rarer molecules, namely isotopologues of water and carbon monoxide. We searched for HDO, \ce{H2^{18}O} and $^{13}$CO. We could only put upper limits for \ce{H2^{18}O} and $^{13}$CO, but our retrieval favour the presence of HDO at 2$\sigma$. We searched for HDO in correlation maps, and we could not see a significant maximum of correlation from this molecule. We included HDO in our H$_2$O models but no significant change in the maximum value in the correlation map is observed. However, the Bayes factor (see Sect. \ref{ssec:Bayes}) for a model that includes HDO compared to a model with water and CO only is 12, indicating a strong preference towards a model that contains HDO. 

The tentative detection of HDO in the atmosphere of HIP~67522~b yields a $\log_{10}(\mathrm{HDO/H_2O}) = -1.4 \pm 0.4$, corresponding to a D/H enrichment factor of $\sim$1000 relative to the protosolar value. The implications of this large enrichment are discussed in section \ref{ssec:HDO-disc}. 

In order to assess the robustness of the HDO detection, we ran injection-recovery tests where we injected our best-fit model in the data at -K$_\mathrm{p}$ and tried to retrieve the presence of HDO with correlation and nested sampling. Our results are very consistent with the results on the real planet: we do not see anything striking out of the correlation map, but the HDO posterior from the nested sampling peaks at the injected value with a low MMR tail associated with lower likelihood (almost similar to Fig.~\ref{figure:marg_P0}). We also calculated the Bayes factor of a retrieval where we searched for both HDO and H$_2$O compared to a retrieval with only H$_2$O, and as in the real data it is around 10. Finally, we injected a model with only water and tried to retrieve both water and HDO. We obtain a flat posterior with a 3$\sigma$ limit on the MMR at 10$^{-4.5}$, and a Bayes factor around one when we include or not HDO in the model. Essentially, our HDO detection is very strong statistically and perfectly consistent with the expectations of injection recovery tests. 

\section{Discussion}
\label{sec:discussions-conclusions}

The exceptionally high detection significance of H$_2$O (20$\sigma$) and CO (5$\sigma$) in HIP~67522~b far exceeds typical values obtained through ground-based infrared high-resolution transmission spectroscopy of hot Jupiters \citep[e.g.,][]{Brogi2016, Birkby2018}. This remarkable signal strength directly reflects the planet's extreme physical properties: its low mass combined with its inflated radius yields an atmospheric scale height comparable to the planetary radius itself, dramatically enhancing the transmission signal. 

The velocity offset of $-2.9 \pm 0.2$~km~s$^{-1}$ from the stellar systemic velocity indicates day-to-night atmospheric circulation, as expected for a highly irradiated planet. The upper limit on equatorial rotation ($<2$~km~s$^{-1}$ at 3$\sigma$) is consistent with tidal locking and excludes Jupiter-like rotation rates, suggesting that tidal dissipation can already have synchronized this young planet's rotation with its orbital period. This is an important result for constraining star-planet tidal interactions.

Our derived C/O ratio of $0.87^{+0.05}_{-0.09}$  (isothermal) and $0.92^{+0.04}_{-0.07}$ (two-temperature) indicates super-solar composition while our metallicity retrieval favours supersolar values or [C+O/H] of $+0.6^{+0.4}_{-0.4}$ (isothermal) and $+1.9^{+0.4}_{-0.4}$ (two-temperature) . Our C/O ratio is in conflict with the work of \citet{Thao2024}, which reported solar C/O ratio.

Note that our retrieved C/O and metallicity do not depend strongly on whether the mass uncertainty is included as a free parameter in the retrieval or is fixed, and we are therefore confident that they represent correct values for our isothermal models no matter the mass. However, our metallicity is strongly dependent on the presence of clouds: the higher the clouds (up to 0.1 mbar where they would hide the molecular signatures), the larger the metallicity. This degeneracy cannot be resolved by our data alone, but we note that there is no statistical preference for the presence of clouds as the Bayes factor is unity compared to a clear atmosphere (see Sect. \ref{ssec:Bayes}). If we can validate an independent estimate of the metallicity and the mass, we will be able to infer whether the atmosphere is cloudy with a super solar metallicity or cloud free with a lower metal content. We also note that the presence of clouds has an effect on the mass but of only 10$\%$ at most (as clouds change the radius by a roughly similar amount), which does not resolve our discrepancy with \citet{Thao2024}. 

\subsection{Mass constraints}
\label{ssec:mass}
The most puzzling result of our paper is the large discrepancy in mass with JWST observations, which disagree at 3$\sigma$, favouring a two times more massive planet in our case. Unfortunately, without performing a joint JWST-CRIRES+ retrieval which is out of the scope of this paper, it is unlikely to solve this issue. We can think of two possible flaws that might affect our results: 1) as we fit our radius to the TESS value, we could have an error if there is a strong optical absorber or a very large super-Rayleigh diffusion that would affect the deduced radius because of a much larger atmosphere than predicted in the visible. That would mean that the actual radius used in the gravity estimation is smaller, leading to a smaller mass for a given gravity. This is unlikely, and would need to affect a significant part of the large TESS wavelength range of integration (600--1100~nm roughly) but not impossible. 2) Our results depend on a normalisation factor, which we cannot confirm the validity a posteriori. If there is a normalisation issue somewhere in our data analysis, that might explain the discrepancy. 

If both these caveats are not the source of discrepancy, then the atmospheric model can be the source of error. For example, the mass estimations from \citet{Thao2024} only use equilibrium chemistry models whereas we ran free retrievals because we didn't want to make assumptions on the thermodynamics of such young objects, and this can be a source of discrepancy in the results. Essentially, with a complex temperature and/or composition profile, we might be able to reconcile CRIRES+ and JWST deduced masses (Chabrol et al. in prep.). 

However, as long as the mass of the planet is unknown, the huge degeneracy that we obtain between mass, radius and reference pressure cannot be solved (see Fig. \ref{figure:marg_P0}). Combining with JWST might reduce the extent of this degeneracy, but it is pretty unlikely that it would actually resolve it. This unknown mostly affects the metallicity which is strongly dependent on the gravity through the scale height of the atmosphere, and can only be poorly constrained so far.

\subsection{Bayes factor}
\label{ssec:Bayes}
In this paper, we have compared several different atmospheric models, with different numbers of free parameters. A usual way to perform model comparisons -- hence decide whether a model is statistically favoured compared to another -- in astrophysics is through the use of Bayesian Information Criterion (BIC) or Akaike Information Criterion (AIC), that are comparisons of the likelihoods of different models penalized by the number of parameters. A model that has a poor likelihood with very few parameters will often be preferred to a model that has a better fit but many more parameters. Such criteria are theoretical approaches to model comparison in the context of studies where the probability are defined relatively to an unknown multiplying factor in the Bayes equation, called the evidence which represents the probability of the data given a model. However, in the context of nested sampling, a measure of the evidence is obtained as an output of the algorithm. We can thus use a more robust criterion for model comparison which is to calculate the evidence ratio \citep[which is a statistical equivalent to Occam's razor, see the detailed discussion in chapter 28 of][]{Mackay2003}, also called the Bayes factor and often denoted $K$:
\begin{equation}
    K = \dfrac{P(D|M_1)}{P(D|M_2)}
\end{equation}
where D are the data,  M$_1$ is a first model and M$_2$ another one. The rule of thumb criterion to favour M$_1$ over M$_2$ is when K>3, with a strong decision towards model 1 if K>10. 

With our data, a model that contains water and CO compared to a model that only contains water leads to a Bayes factor of $e^{22}$, which excludes with certainty the absence of CO in the atmosphere. Regarding the temperature structure, the two-temperature model with a free pressure transition is only slightly favoured over the isothermal model with a Bayes factor of 3. However, when fixing the pressure transition to the best retrieved value of 3~mbar, the Bayes factor reaches 20 in favour of the two-temperature model, indicating a strong statistical preference for a non-isothermal atmosphere. The HDO is also strongly favoured when using the evidence ratio.

\subsection{HDO/H$_2$O ratio: signature of strong escape?}
\label{ssec:HDO-disc}

The tentative 2$\sigma$ detection of HDO would represent the first such measurement in an exoplanet atmosphere. It is associated with a Bayes factor of 12, which indicates a strong statistical preference towards the presence of this molecule in the atmosphere. However, if it is real, we must be able to explain the D/H enrichment factor of $\sim$1000 relative to the protosolar ratio. This enrichment is unlikely to arise from formation processes \citep[and private comm. with C. Morley]{Morley2019}: such extreme deuterium fractionation largely exceeds the primordial ratio measured in protoplanetary disks \citep[$\sim 2 \times 10^{-3}$;][]{Tobin2023}. 

The only other known D/H ratio of such extreme value is in Venus's atmosphere \citep[$\log(\mathrm{HDO/H_2O}) \approx -1.5$;][]{Donahue1982}, where $\sim$4.5~Gyr of preferential hydrogen escape produced a $\sim$100-fold enrichment relative to Earth. 
Notably, recent Venus Express analysis revealed that the venusian D/H ratio increases from $\sim$160$\times$ terrestrial at the cloud tops to $\sim$1500$\times$ at 108~km altitude \citep{Mahieux2024}, demonstrating that upper atmospheric layers can be significantly more enriched than the bulk atmosphere. As high-resolution transmission spectroscopy probes the upper atmosphere where photochemical fractionation and escape occur, our measurement may therefore reflect local fractionation in the escaping atmosphere, consistent with predictions that HDO could trace atmospheric evolution in irradiated exoplanets \citep{Morley2019, molliere2019}.

 For the 17~Myr old HIP~67522~b to achieve comparable enrichment than Venus requires extraordinarily rapid escape rates, plausibly driven by the intense XUV flux from its magnetically active, frequently flaring host star \citep{Ilin2025a}. Enhanced D/H ratios in exoplanet atmospheres due to atmospheric escape are predicted by numerical models \citep[e.g.][]{Cherubim2024}. The D/H ratio provides constraints on the atmospheric escape mechanism, particularly if the He/H ratio is measured as well, which allows to distinguish between competing escape mechanisms.  Follow-up observations at higher signal-to-noise and different wavelength ranges are essential to confirm this detection, and to investigate further whether young, highly irradiated planets routinely exhibit signatures of ongoing deuterium fractionation. Observations in the Y-band near the He triplet at 1083~nm will allow to eventually measure He/H if excess Helium absorption is measured in the exoplanet atmosphere. Additionally, as pointed out by \citet{Cherubim2024}, deuterium-enhanced planets should be helium-dominated and methane-depleted, meaning that there is a favourable wavelength range to detect HDO at 3.7~$\mu$m, a wavelength range accessible with few high-resolution spectrographs including CRIRES+, iSHELL \citep{Rayner2022}, and the future Metis instrument at the Extremely Large Telescope \citep{Brandl2021}.

\section*{Data availability}
The reduced data are available on Zenodo\footnote{\url{https://doi.org/10.5281/zenodo.17909093}} \citep{hip67522data}. The raw data as well as the ESO-processed data are all publicly available one year after the observations at the ESO archive. The planet isothermal synthetic spectra (as well as associated information about opacity sources) are also available at the Zenodo record containing the reduced data.

\begin{acknowledgements}
Based on observations made with ESO Telescopes at the La Silla Paranal Observatory under programme ID 114.28HP. We are grateful to the ESO Director General and Director of Science for granting us DDT for this project, to the ESO observing staff, and to Michael Way and Caroline Morley for helpful discussions. 
F. Debras acknowledges funding from the French National Research Agency (ANR) project ExoATMO (ANR-25-CE49-6598). F.Debras would like to warmly thank the \textit{Action Thématique Physique Stellaire} (ATPS) and \textit{Action Thématique ExoSystèmes} (AT-EXOS) of CNRS/INSU co-funded by CEA and CNES for their financial support.
B. Klein., S. Aigrain acknowledge funding from the European Research Council under the European Union’s Horizon 2020 research and innovation programme (grant agreement No 865624, GPRV).
A. Masson acknowledges support and funds from the grants nº CNS2023-144309 by the Spain Ministry of Science, Innovation and Universities MICIU/AEI/10.13039/501100011033. T. Hood acknowledges funding from the French National Research Agency (ANR) project EXOWINDS (ANR-23-CE31-0001-01).

This work was granted access to the HPC resources of CALMIP supercomputing center under the allocation 2021-p21021.
This work made use of Astropy: a community-developed core Python package and an ecosystem of tools and resources for astronomy \citep{astropy:2013, astropy:2018, astropy:2022}, numpy \citep{Harris2020}. This research has made use of the Astrophysics Data System, funded by NASA under Cooperative Agreement 80NSSC21M0056. This research has made use of the SIMBAD database, operated at CDS, Strasbourg, France. The data reduction process was sped up thanks to \texttt{GNU parallel} \citep{tange_2023_10199085}. 
\end{acknowledgements}

\begin{figure*}
\centering
\includegraphics[width=0.7\textwidth]{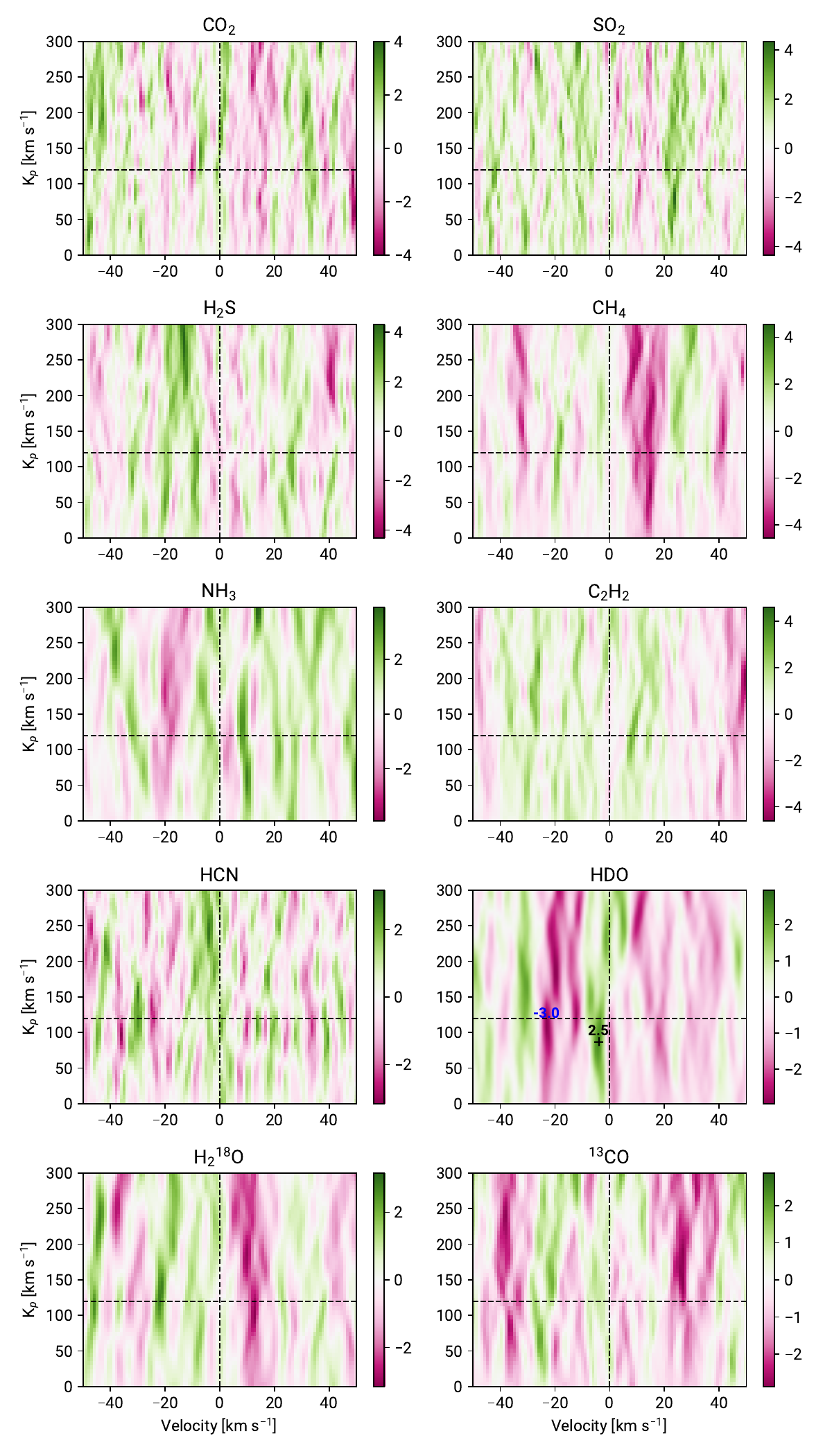}
\caption{Same as Fig.~\ref{figure:phase-vsys} for all the non-detected species.}
\label{figure:kp-vsys-nodetection}
\end{figure*}

\bibliographystyle{aa} 
\bibliography{hip} 

\begin{appendix}
\section{Posterior distribution for the two-temperature models}
\begin{figure*}
\centering
\includegraphics[width=0.7\textwidth]{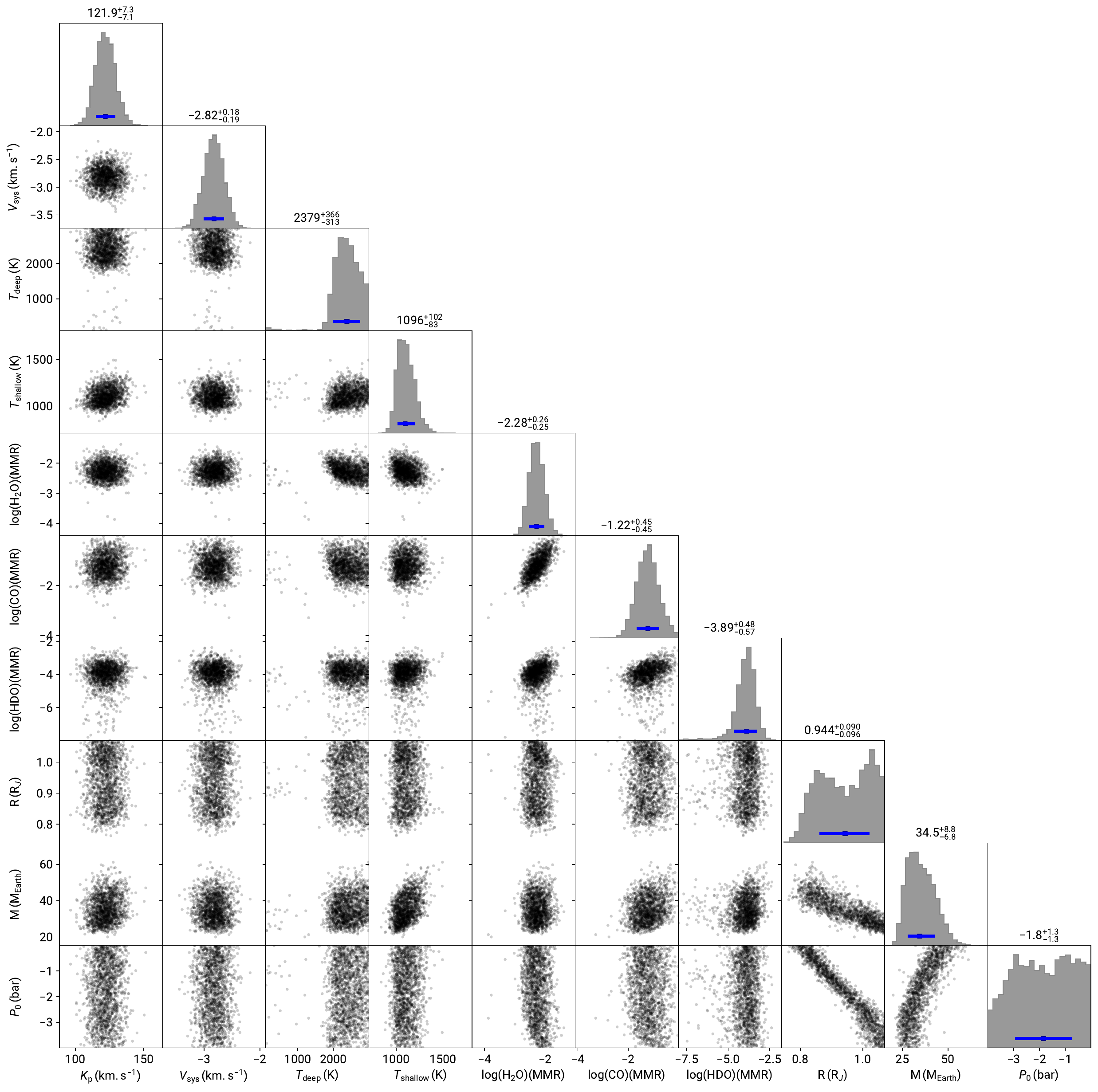}
\caption{Corner plot of the posterior distribution of our two-temperature models.}
\label{figure:marg2T}
\end{figure*}

\end{appendix}
\end{document}